\documentstyle[12pt,epsf,axodraw]{article}
\def\tom{\tau^- \rightarrow \omega \pi^- \nu_{\tau}}
\def\tphi{\tau^- \rightarrow \phi\pi^- \nu_{\tau}}
\def\be{\begin{equation}}
\def\ee{\end{equation}}

\begin{document}
\vspace*{.3cm}
\begin{flushright}
\large{CINVESTAV-FIS-09/96}
\end{flushright}
\begin{center}
\LARGE{\bf VMD description of $\tau \rightarrow (\omega,\ \phi) \pi^- 
\nu_{\tau}$ decays and the $\omega-\phi$ mixing angle}
 \end{center}
\vspace{.8cm}
\begin{center}
\Large G. L\'opez Castro and D. A. L\'opez Falc\'on\\
\vspace*{.4cm}
{\normalsize{\it  Departamento de F\'\i sica, Cinvestav del IPN, Apdo. \\ 
\vspace{-.4cm} \it  Postal 14-740, 07000 M\'exico, D.F., MEXICO.}}
\vspace*{.4cm}
\end{center}

\thispagestyle{empty}
\centerline{ \bf Abstract}
\vspace{.3cm}
  Using the vector meson dominance model we get predictions for the 
Cabibbo-favored $\tom$ and $\tphi$ decays. We show how the measurements of 
these two decays can provide information on the nature of the violation of 
the OZI rule.

 \vspace{.5cm} 

\newpage
\setcounter{page}{1}
\vspace{2cm}

   $\tom$ and $\tphi$ are two Cabibbo-favored, marginal decay modes of 
the $\tau$ lepton whose precise measurements can give insight into the 
existence of axial second class currents [1] and 
 on the nature of the violations to the OZI rule [2] in a clean 
environment, respectively.

  The decay $\tom$ has been studied previously in the context of the 
vector-meson dominance model [3], a low energy $U(3)\times U(3)$ chiral 
lagrangian model [4] and, most recently, using the heavy vector-meson 
chiral perturbation formalism [5]. The experimental information about 
this decay involves the measurement of the decay rate and the spectral 
function [6]. The angular distribution of the $\omega \pi^-$ system was 
found to be consistent with a $J^P =1^-$ state [6], which is typical of 
first class vector current and is in agreement [7] with data on $e^+e^- 
\rightarrow \omega \pi^0$ via the Conserved Vector Current (CVC) hypothesis.

  Since the $\phi(1020)$ is almost a pure $\bar{s} s$ state, the decay 
$\tphi$ is expected to be very suppressed with respect to $\tom$ due to 
the OZI rule. 
Actually, the experimental upper limit on this decay indicates that 
$B(\tphi) < 2.6 \times 10^{-4}$ at the 90 \% C.L. [8].  Theoretical 
estimations based on the CVC hypothesis and using the $e^+e^- 
\rightarrow \phi \pi^0$ data give $B(\tphi) < 9 \times 10^{-4}$ [7]. Thus, 
the measurement of the $\phi \pi^-$ production relative to $\omega\pi^-$ 
in $\tau$ decays can provide clean information on the nature of the 
violations to the OZI rule.

 In this paper we examine these two decays of the $\tau$ lepton by using 
the model proposed in Ref. [3]. We first extract the relevant coupling 
constants from light meson decays and predict $B(\tom) = (1.22 \pm 0.56)\%$ 
and 
$B(\tphi) = (1.20 \pm 0.48) \times 10^{-5}$. Then we estimate the ratio of 
these two 
decays and find it to be sensitive to the $\omega-\phi$ mixing angle.

  The lowest order amplitude for the $\tau^- \rightarrow V \pi^- \nu_{\tau}$ 
decays can be written as 
\be
{\cal M} = \frac{G_F V_{ud}}{\sqrt{2}} l^{\mu} < V\pi^- | \bar{u} 
\gamma_{\mu} d | 0>,
\ee
where $l^{\mu}$ is the V--A leptonic current and $V_{ud}$ is the $ud$ 
element of the Cabibbo-Kobayashi-Maskawa matrix. The form of the hadronic 
matrix element of the vector current is fixed by Lorentz covariance to be:
\be
< V\pi | \bar{u} \gamma_{\mu} d | 0> = i F_V(s) \epsilon_{\mu \alpha 
\beta \gamma} \varepsilon^{*\alpha} q_1^{\beta} q_2^{\gamma}
\ee
where $q_{1,2}$ denote the four-momenta of $V$ and $\pi^-$; $\varepsilon^*$ 
is the polarization four-vector of $V$, and $F_V(s)$ is the $s$-dependent 
($s=(q_1+q_2)^2$) form factor for the hadronic vertex.

  The squared mass distribution of the hadronic system, can be written as 
follows,
\be
\frac{d\Gamma}{ds} = \frac{G_F^2 |V_{ud}|^2}{ 1536 \pi^3 m^3} 
|F_V(s)|^2 \frac{(m^2-s)^2}{s^2} (m^2 + 2s) (s^2-2s\Sigma^2 +\Delta^4)^{3/2}
 \ee
where $m$ denotes the $\tau$ mass, $\Sigma^2\equiv m_V^2+m_{\pi}^2$ and 
$\Delta^2\equiv m_V^2-m_{\pi}^2$. The last factor in the r.h.s of Eq. (3) is 
characteristic of the $p$-wave of the $V\pi^-$ system. Eq. (3) above is in 
agreement with Eq. (5) of Ref. [3], after including  a factor 1/2 missing in 
their Eq. (5).

  The spectral function $v(s)$ for this decay is defined as follows [6],
\be
v(s) = \frac{32 \pi^2 m^3}{G_F^2 |V_{ud}|^2(m^2-s)^2(m^2+2s)}\ 
\frac{d\Gamma}{ds}.
\ee
  In order to make predictions we need a specific model for $F_V(s)$. In 
the vector-meson dominance model, this form factor is given by (See Fig. 1)
\be
F_V(s) =  \Sigma_{V'=\rho,...} \frac{g_{V'} g_{V'V 
\pi^-}}{m_{V'}^2-s-im_{V'} \Gamma_{V'}}
\ee
where $g_{V'}$ denotes the coupling constant describing the $W^--V'$ 
junction and $g_{V'V\pi^-}$ is the $V'V\pi^-$ coupling constant.

  As is shown in Ref. [6], the experimental data on the spectral function 
requires the presence of at least one $\rho'$ in addition to the 
$\rho(770)$. Including these two resonances ($\rho + \rho'$), the form 
factor can be written as follows:
\be
F_V(s) = \frac{g_{\rho^-}g_{\rho V 
\pi^-}}{m_{\rho}^2-s-im_{\rho}\Gamma_{\rho}} \left\{ 1 + \alpha_V 
\frac{m_{\rho}^2-s-im_{\rho}\Gamma_{\rho}}{m_{\rho'}^2 
-s-im_{\rho'}\Gamma_{\rho'}} \right\}
\ee
where $\alpha_V \equiv g_{\rho'}g_{\rho' V \pi}/g_{\rho}g_{\rho V \pi}$.

  Thus, the data on the spectral function for $\tom$ can be used as a 
test of the form 
factor model. In order to fit the data on $v(s)$, we have used in Eq. (6) 
the values $g_{\rho^-}=\sqrt{2}m_{\rho}^2/\gamma_{\rho} = (0.166\pm0.005)\ 
{\rm GeV}^2$ [3, 9], and $g_{\rho^+ \omega \pi^-}=(13.5 \pm 2.5)\ {\rm 
GeV}^{-1}$, 
which cover the range of values for this coupling as extracted from 
$\rho^{-,\ 0} 
\rightarrow \pi^{-,\ 0} \gamma,\ \omega \rightarrow \pi^0 \gamma$ and $\pi^0 
\rightarrow \gamma \gamma$ using the isospin symmetry relation $g_{\rho^- 
\omega \pi^-} = g_{\rho^0 \omega \pi^0}$. We have fitted the data of 
Ref. [6] using 
the $\rho(1450)$ or the $\rho(1700)$ in addition to the $\rho(770)$, 
leaving $\alpha_{\omega}$ as a free parameter and allowing for an overall 
normalization factor $N$ of the spectral function.

  Although the data are rather poor, we find that a better fit can be 
obtained using the $\rho +\rho(1450)$ combination. In this case the best fit 
gives: \begin{eqnarray}
\alpha_{\omega} &=& -0.37 \pm 0.14\\
N &=& 1.22 \pm 0.34\ . 
\end{eqnarray}
  Although the normalization factor is consistent with unity, the 
experimental data for the spectral function could be systematically a 20 \% 
higher than expected.
 Let us mention that the effects of taking an $s$-dependent width [3] for the 
$\rho$ and $\rho'$  Breit-Wigner forms in Eq. (6) does not change the 
results of the fit.

   Using the above results into Eqs. (6) and (3) we obtain (for $N=1$):
\be
B(\tom) = (1.22 \pm 0.56)\%
\ee
where the uncertainty reflects the errors in $g_{\rho \omega \pi},\  
g_{\rho}$ and $\alpha_{\omega}$. The large uncertainty above is 
dominated by the error we have attributed to the $\rho\omega\pi$ 
coupling constant. This result for the branching ratio is 
consistent (within errors) with the experimental value $B(\tom) = (1.6 \pm 
0.5) \%$ [10] but it lies below the prediction based on the CVC hypothesis, 
$B^{CVC}(\tom) =(2.2\pm 0.3)\% $ [7].

  Since the CVC hypothesis is expected to be exact in the limit of 
isospin symmetry, one should expect small deviations from the result of 
Ref. [7]. There are two probably reasons for this discrepancy. On the one 
hand, the cross section data on $e^+e^- \rightarrow \omega\pi^0$ [11] 
used to get the CVC prediction for $\tom$ was measured only for $1\ {\rm 
GeV} \leq E_{cm} \leq 1.4\ {\rm GeV}$, and an extrapolation is required 
for the kinematical range in $\tau$ decay. Actually, the model used in 
[7] to extrapolate the $e^+e^-$ data seems to be in disagreement with the 
measurement of the spectral function in $\tom$ at higher $s$ values (see 
Ref. [6]). This 
suggests that the errors in the CVC prediction for $\tom$ could have been 
underestimated. On the other hand, let us comment that if the 
normalization in the data for the spectral function reported in [6] are 
indeed  larger by 20 \% (see Eq. (8)), this would increase our prediction 
by the same amount.

   In order to predict the branching ratio for $\tphi$, we can rely on the 
above model and use the $\rho^+ \phi \pi^-$ coupling extracted from $\phi 
\rightarrow \rho \pi$:
 \be
g_{\rho^+ \phi \pi^-} = (1.10 \pm 0.03)\ {\rm GeV}^{-1}.
\ee
Observe that we have to divide the $\phi \rightarrow \rho \pi$ decay rate 
reported in [10] by a factor of  
3 in order to account for the three isospin channels allowed in $\phi 
\rightarrow \rho \pi$ decay.

  Since the spectral function of $\tphi$ has not been measured yet, we 
will assume that $\alpha_{\phi} = 
\alpha_{\omega}$ in Eq. (6), which is equivalent to require that:
\be
\frac{g_{\rho \phi \pi}}{g_{\rho \omega \pi}}
=\frac{g_{\rho' \phi \pi}}{g_{\rho' \omega \pi}}.
\ee
This relation can be obtained if we assume a $U(3)$ invariant 
coupling for the $V'VP$ vertex and we replace the $1^3S_1$ nonet of 
vector mesons (the $\rho$) by the $2^3S_1$ nonet of vector mesons 
(the $\rho'$) (in 
the $n^{2S+1}L_J$ spectroscopic notation. See for example p.1320 in [10]).

   Using Eq. (10) into (6) and (3) and relying on the above assumptions, we 
obtain: \be
B(\tphi) = (1.20 \pm 0.48) \times 10^{-5}
\ee
which lies one order of magnitude below the experimental upper limit 
reported in Ref. [8], 
and almost two order of magnitude below the upper limit obtained using the 
CVC hypothesis 
[7]. The error in Eq. (12) is dominated by the error in $\alpha_{\phi}$.

  We now focus our discussion on the $\omega-\phi$ mixing angle. As is 
well known (see p.1320 in Ref. [10]),  the physical states $\omega$ and 
$\phi$, can be written 
in terms of the octet ($V_8$) and singlet ($V_0$) states of SU(3) as follows:
\begin{eqnarray}
\omega &=& V_8 \sin \theta_V + V_0 \cos\theta_V \nonumber \\
\phi &=& V_8 \cos\theta_V - V_0\sin\theta_V
\end{eqnarray}
or, if we define $\delta \equiv \theta_V -\theta_I$, which measures the 
deviation 
from the ideal mixing angle $\theta_I = \arctan(1/\sqrt{2}) \approx 
35.3^o$, we can write:
\begin{eqnarray}
\omega&=&\cos\delta \frac{1}{\sqrt{2}}(\bar uu+\bar dd) -\sin\delta 
\bar ss \nonumber \\
\phi&=&-\sin\delta \frac{1}{\sqrt{2}}(\bar uu+\bar dd) - \cos\delta \bar ss.
\end{eqnarray}
The value obtained from the quadratic mass formula for vector mesons is 
$\delta \approx 4^o$ (p. 1320 in [10]).

   The SU(3) invariant lagrangian for the $V'VP$ interactions including 
singlet vector states can be written as follows:
\begin{eqnarray}
{\cal L}_{V'VP} &=&
 G_{V'VP}^8 d_{abc} \epsilon^{\alpha \beta \gamma 
\delta} P^a \partial_{\alpha}V_{\beta}^b \partial_{\gamma}{V'}_{\delta}^c 
\nonumber \\
&& \ + \sqrt{\frac{2}{3}}G_{V'VP}^0 \delta_{ab} \epsilon^{\alpha \beta 
\gamma 
\delta} P^a (\partial_{\alpha}V_{\beta}^b \partial_{\gamma}{V'}_{\delta}^0+ 
V\leftrightarrow V')
 \end{eqnarray}
where $P^a(V^a)\, a=1,\cdots,8$ denote the octet of pseudoscalar (vector) 
mesons, and $V^0$ is the SU(3) singlet vector meson. $G_{V'VP}^i\ (i=8,\ 
0)$ are the corresponding octet and singlet coupling constants.

  From the above lagrangian and using the definition given in Eqs. 
(13)--(14) for the physical states we can derive
\begin{eqnarray}
g_{\rho^+ \phi \pi^-} &=& \frac{G^8_{V'VP}}{3} \left\{ 
\sqrt{2}(1-r)\cos\delta -(1+2r)\sin\delta \right\} \\
g_{\rho^+ \omega \pi^-} &=& \frac{G^8_{V'VP}}{3} \left\{ 
\sqrt{2}(1-r)\sin\delta +(1+2r)\cos\delta \right\} 
\end{eqnarray}
where $r\equiv G^0_{V'VP}/G^8_{V'VP}\neq 1$ accounts for deviations from 
nonet symmetry of vector mesons.

  If we introduce now these couplings into Eq. (6) and use the results 
given in Eqs. (9) and (12) we can build the following ratio:
\begin{eqnarray}
R_{\omega\phi} &\equiv& \frac{B(\tphi)}{B(\tom)} \nonumber \\
&\approx& 0.1482 \left | \frac{\sqrt{2} 
\left(\frac{1-r}{1+2r}\right)-\tan\delta}{\sqrt{2} 
\left( \frac{1-r}{1+2r}\right) \tan\delta + 1} \right |^2
\end{eqnarray}
Note that the above result reduces to the simple model proposed in Ref. 
[12] (namely, that violations to the OZI rule arises purely from 
$\omega-\phi$ mixing) in the limit that $r=1$, because in this case \be
R_{\omega\phi} = \tan^2\delta \cdot f
\ee
where $f$ is a kinematical factor.

  In Table 1 we show the results for $R_{\omega\phi}$, Eq. (18), as a 
function of 
the angle $\delta$ when we allow for a $\pm 20\%$ deviation from nonet 
symmetry ($r=1$). We can observe that, when $r=0.8\ {\rm or}\ 1$, the 
relative production of 
$\phi\pi^-/\omega \pi^-$ in tau decays is very sensitive to deviations 
from the ideal mixing angle. In the special case that $\delta = 4^0$ and 
$r=1$, and using the experimental branching ratio for $\tom$, we obtain from 
Table 1 $B(\tphi) \approx 1.16 \times 10^{-5}$, which is consistent with 
the estimated value in Eq. (12).

In summary, we have given a description of the $\tau^- \rightarrow (\omega,\ 
\phi) \pi^- \nu_{\tau}$ in the framework of the vector dominance model. 
The branching fraction of $\tom$ is found to be consistent with the 
present 
experimental value [10] and it lies below the prediction based on the CVC 
hypothesis. Our prediction for the $\phi \pi^-$ mode $B(\tphi) = 
(1.20 \pm 0.48) \times 10^{-5}$ lies one order of magnitude below the 
present experimental upper limit given in 
Ref. [8] and could be measured at a $\tau$-charm factory. It is also 
shown that the simultaneous measurements of the $\omega \pi^-$ and $\phi 
\pi^-$ decay rates, can provide useful information on the $\omega-\phi$ 
mixing angle.
 
\

{\bf Acknowledgements}

  One of the authors (GLC) is grateful to Dr. Roger Decker for useful 
discussions on the subject of this paper. We would like to dedicate 
this paper to his memory.

\newpage

\

\begin{center}
TABLE CAPTIONS
\end{center}

\begin{enumerate}
\item Predictions for $R_{\omega\phi}$, Eq. (18), as a function of the 
mixing angle $\delta$ for three different choices of $r$.
 \end{enumerate}

\

\begin{center}
\begin{tabular}{|c|c|c|c|}
\hline
$\delta (^\circ)$&\multicolumn{3}{c|}{$R_{\omega\phi}$}\\
\cline{2-4}
&$r=0.8$&$r=1.0$&$r=1.2$\\
\hline
$1$&$1.23\times 10^{-3}$&$4.52\times 10^{-5}$&$1.51\times 10^{-3}$\\
$2$&$8.02\times 10^{-4}$&$1.81\times 10^{-4}$&$2.08\times 10^{-3}$\\
$3$&$4.66\times 10^{-4}$&$4.07\times 10^{-4}$&$2.75\times 10^{-3}$\\
$4$&$2.20\times 10^{-4}$&$7.25\times 10^{-4}$&$3.52\times 10^{-3}$\\
$5$&$6.60\times 10^{-5}$&$1.13\times 10^{-3}$&$4.38\times 10^{-3}$\\
\hline
\end{tabular}
\end{center}
\begin{center}
Table 1
\end{center}

\newpage
\begin{center}
\begin{picture}(324,120)(0,0)
\Text(10,69)[1]{$\tau^-$}
\ArrowLine(22,69)(50,69) \Vertex(50,69){2}
\ArrowLine(50,69)(78,97) \Text(84,103)[lb]{$\nu_\tau$}
\Photon(50,69)(78,41){2}{4} \Text(64,55)[rt]{$W^-$}
\Line(78,41)(106,13) \Text(112,7)[lt]{$\pi^-$}
\Line(78,41)(106,69) \Text(112,75)[lb]{$V$}
\GCirc(78,41){5}{.5}
\Text(160,69)[c]{$\displaystyle =\sum_{V'=\rho,\,\rho^{'},\ldots}$}
\Text(214,69)[1]{$\tau^-$}
\ArrowLine(226,69)(254,69) \Vertex(254,69){2}
\ArrowLine(254,69)(282,97) \Text(288,103)[lb]{$\nu_\tau$}
\Photon(254,69)(268,55){2}{4} \Text(262,62)[rt]{$W^-$}
\Vertex(268,55){2}
\Line(268,55)(282,41) \Text(275,50)[lb]{V$'$}
\Vertex(282,41){2}
\Line(282,41)(310,13) \Text(316,7)[lt]{$\pi^-$}
\Line(282,41)(310,69) \Text(316,75)[lb]{$V$}
\end{picture}
\end{center}

\

\begin{center}
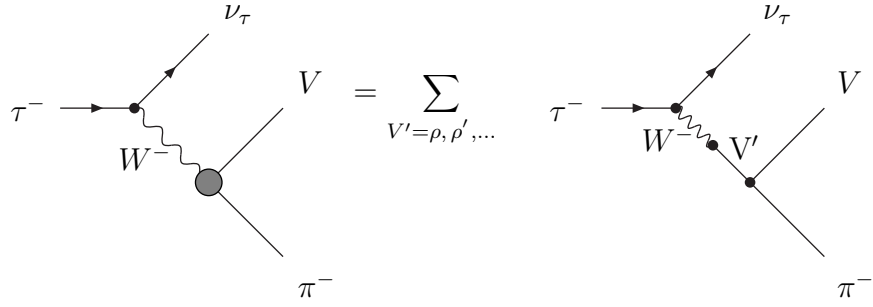
Fig. 1: VMD contributions to $\tau^-\rightarrow V\pi^-\nu_\tau$.
\end{center}


\begin{thebibliography}{99}
\bibitem{ref1}
S. Weinberg, Phys. Rev. {\bf 112}, 1375 (1958); C. Leroy and J. Pestieau, 
Phys. Lett. {\bf B72}, 398 (1978).
\bibitem{ref2} 
S. Okubo, Phys. Lett. {\bf 5}, 165 (1963); G. Zweig, CERN Report no. 
8419/TH, 412 (1964); J. Iizuka, Prog. Theor. Phys. Suppl. {\bf 
37-8}, 21 (1966).
\bibitem{ref3}
R. Decker, Z. Phys. {\bf C36}, 487 (1987).
\bibitem{ref4}
S. Fajfer, K. Suruliz and R. J. Oakes, Phys. Rev. {\bf D46}, 1195 (1992).
\bibitem{ref5} 
H. Davoudiasl and M. B. Wise, Phys. Rev. {\bf D53}, 2523 (1996).
 \bibitem{ref6}
H. Albrecht {\em et al.}, ARGUS Collaboration, Phys. Lett. {\bf B185}, 
223 (1987).
 \bibitem{ref7}
S. I. Eidelman and V. N. Ivanchenko, Phys. Lett. {\bf B257}, 437 (1991).
\bibitem{ref8}
M. Daoudi, CLEO Collaboration, Talk presented at the 5th. Int. Symposium 
on Heavy Flavor Physics, Montreal, July 6-10 (1993); J. P. Alexander, 
CLEO Collaboration, ``{\em Search for $\tphi$}", Contributed paper to the 
XVI Lepton-Photon Symposium, Cornell University, (1993).
\bibitem{ref9}
See for example: L. Okun, {\sl Lepton and quarks}, (North Holland, 
Amsterdam, 1982), p. 109.
\bibitem{ref10}
L. Montanet {\em et al.}, {\sl Review of Particle Properties}, Phys. Rev. 
{\bf D50} Part I, (1994).
 \bibitem{ref11}
S. I. Dolinsky {\em et al.}, Phys. Lett. {\bf B174}, 453 (1986).
\bibitem{ref12}
H. J. Lipkin, Phys. Lett. {\bf B60}, 371 (1976).
\end{thebibliography}
\end{document}